\documentclass[pdflatex,sn-mathphys-num]{sn-jnl}

\usepackage{graphicx}%
\usepackage{multirow}%
\usepackage{amsmath,amssymb,amsfonts}%
\usepackage{amsthm}%
\usepackage{mathrsfs}%
\usepackage[title]{appendix}%
\usepackage{xcolor}%
\usepackage{textcomp}%
\usepackage{manyfoot}%
\usepackage{booktabs}%
\usepackage{algorithm}%
\usepackage{algorithmicx}%
\usepackage{algpseudocode}%
\usepackage{listings}%
\usepackage{braket}%
\usepackage{quantikz}%

\newcounter{hypothesis}
\renewcommand{\thehypothesis}{\arabic{hypothesis}}

\newenvironment{hypothesis}[1][]{
 ~\refstepcounter{hypothesis}\vspace{1em} %
  \noindent\textbf{Hypothesis~\thehypothesis.}~#1
}{
  \vspace{1em} 
}

\theoremstyle{thmstyleone}

\theoremstyle{thmstyletwo}%

\theoremstyle{thmstylethree}%

\begin{document}

\title[Article Title]{Mid-circuit measurement as an algorithmic primitive}

\author*[1]{\fnm{Antoine} \sur{Lemelin}}\email{antoine.lemelin.2@ens.etsmtl.ca}

\author[1,2]{\fnm{Christophe} \sur{Pere}}

\author[1]{\fnm{Olivier} \sur{Landon-Cardinal}}

\author[1]{\fnm{Camille} \sur{Coti}}

\affil*[1]{\orgname{École de technologie supérieure – Université du Québec}, \orgaddress{\street{1100 R. Notre-Dame O}, \city{Montréal}, \postcode{H3C 1K3}, \state{Québec}, \country{Canada}}}

\affil[2]{\orgname{PINQ$^2$}, \orgaddress{\street{1950 rue Roy}, \city{Sherbrooke}, \postcode{J1K 1B7}, \state{Québec}, \country{Canada}}}



\abstract{
    We explore the usefulness of mid-circuit measurements to enhance quantum algorithmics. Specifically, we assess how quantum phase estimation (QPE) and mid-circuit measurements can improve the performance of variational quantum algorithms. Our focus is on the single-qubit version of QPE, namely the Hadamard test, applied to the evolution operator. We demonstrate that a mid-circuit measurement acts as a low-energy filter when the desired outcome is obtained. When the undesired outcome is measured, we heuristically rely on a unitary mixer to repopulate low-energy states. Numerical simulations show that this method effectively amplifies the ground state occupancy. We validate our approach on real quantum hardware, namely the IBM Quantum System One \texttt{ibm\_quebec}. }

\maketitle
\pagebreak

\section{Introduction}\label{sec1}
Variational quantum algorithms (VQAs) such as the Quantum Approximate Optimization Algorithm (QAOA)~\cite{Farhi_Goldstone_Gutmann_2014} and the Variational Quantum Eigensolver (VQE)~\cite{Peruzzo_McClean_Shadbolt_Yung_Zhou_Love_Aspuru-Guzik_O’Brien_2014} have emerged as some of the most promising candidates for achieving quantum advantage on near-term quantum hardware. These hybrid algorithms harness both quantum and classical computational resources by relying on a classical optimizer to iteratively adjust the parameters of a quantum circuit based on measurement outcomes.

However, this hybrid structure comes at a cost. The frequent back-and-forth exchange of information between the quantum and classical systems is not only resource-intensive but also inherently limited in its ability to fully explore the quantum Hilbert space. Moreover, VQAs often suffer from phenomena such as barren plateaus and optimizer inefficiencies, which can severely hinder convergence and scalability~\cite{Larocca_Thanasilp_Wang_Sharma_Biamonte_Coles_Cincio_McClean_Holmes_Cerezo_2025, McClean_Boixo_Smelyanskiy_Babbush_Neven_2018}. These issues stem from treating the quantum circuit as a black box and relying solely on classical feedback, without leveraging the internal structure and dynamics of the quantum system itself.

Mid-circuit measurements have long been a staple in quantum error correction schemes. They are prominently featured in codes such as Shor’s nine-qubit code and surface codes, where intermediate measurements are essential for detecting and correcting errors without collapsing the quantum state entirely~\cite{Kitaev_1997,Shor_1995}. The utility and widespread adoption of mid-circuit measurements in these foundational protocols helped establish their importance in quantum computing. While their use was historically confined to error correction, their broader algorithmic potential began to attract interest. As early as the mid-1990s and early 2000s, researchers explored their role in the Quantum Fourier Transform (QFT) and Shor’s algorithm~\cite{Giorda_Iorio_Sen_Sen_2004,Griffiths_Niu_1996,Shor_1997}. The idea remained largely dormant for years, receiving little attention outside the realm of error correction, until IBM enabled support for mid-circuit measurements on real quantum hardware in 2021~\cite{Corcoles_Takita_Inoue_Lekuch_Minev_Chow_Gambetta_2021}, reigniting interest in leveraging this feature for algorithm design. This has led to new proposals that integrate mid-circuit measurements into algorithmic primitives such as the QFT~\cite{Bäumer_Tripathi_Seif_Lidar_Wang_2024}, paving the way for more efficient and potentially fully quantum alternatives to hybrid approaches.

In this work, we introduce a novel quantum-native approach that circumvents the need for a classical optimizer. Our method is based on mid-circuit measurements and conditional feed-forward, forming a purely quantum feedback loop entirely. The algorithm uses the binary outcome of a single-qubit Quantum Phase Estimation (QPE) procedure~\cite{Kitaev_1995} to guide the evolution of the system toward low-energy states. 

We show mathematically that when the ancilla qubit yields the desired outcome, the quantum state is biased towards the ground space in a coherent and systematic way. When the undesired outcome is obtained, applying a mixer operation heuristically repopulates the low-energy states.

Numerically, we observe that this feedback-driven evolution exhibits convergence toward optimal solutions. We propose a practical implementation of this approach and show that, across iterations, the system consistently amplifies high-quality solutions with increasing fidelity. This result suggests that mid-circuit measurements, when paired with targeted interference control, can serve as a powerful tool to enhance quantum optimization performance and move beyond the limitations of current variational strategies.

\section{Methods}
\subsection{Amplifying Ground State Amplitude via one qubit Quantum phase estimation}
To investigate how mid-circuit measurements and interference can be used to favor low-energy eigenstates, we used a variant of Quantum Phase Estimation (QPE)~\cite{Kitaev_1995} using a single ancilla qubit, a technique often referred to as the Hadamard test~\cite{Cleve_Ekert_Macchiavello_Mosca_1998}. The use of a single ancilla rather than a many qubit quantum register is dictated by current quantum hardware and has been the object of recent proposals~\cite{Faehrmann_Eisert_Kueng_2025}. This approach allows us to examine whether our quantum procedure systematically amplifies the ground-state component of a given Hamiltonian or can lead to a dead end. Related ideas appeared in prior work~\cite{Griffiths_Niu_1996,Rall_2021}, leading to methods typically relying on applying a unitary operator $U$ a number of times that grows exponentially with precision $2^{\frac{1}{\epsilon}}$ where $\epsilon$ is . In contrast, our objective is not to estimate eigenvalues or amplitudes, but to amplify the ground state amplitude through quantum interference. It requires only a single application of a quantum circuit per iteration at the price of a larger circuit depth. This means that each iteration requires only one run of the circuit, though the depth of that circuit increases compared to repetition-based schemes.

Let's now describe our approach in details. First, we consider a Hamiltonian $H$ which is diagonal in the computational basis. This is a strong assumption for our method to work. We will emphasize these assumptions by enumerating them throughout the article:

\begin{hypothesis}
The eigenbasis of the Hamiltonian $H$ is the computational basis. 
\end{hypothesis}

In the context of optimization, the ground energy level $E_0$ corresponds to the optimal solution (i.e., the one with the lowest cost), while $E_{\max}$ represents the worst possible solution. To ensure that the optimal solution corresponds to the ground state of the system, it is possible to construct a Quadratic Unconstrained Binary Optimization (QUBO) formulation in which the minimum eigenvalue of the associated Hamiltonian is exactly zero. This is achieved by assigning positive penalties to constraint violations, such that the energy (or cost) becomes zero only when all constraints are satisfied. A concrete example is the \textit{Vertex Cover} problem, which we will detail in Section~\ref{subsec:problem_definition}. 

\begin{figure}[htbp]
    \centering
    \begin{quantikz}
    \lstick{$\ket{0}$}&\gate{H}&\ctrl{1}&\gate{R_Z(\theta)}&\gate{H}&\meter{}\\
    \lstick{$\ket{\psi}$} & \qw //^n & \gate[1]{e^{-iHt}} & \qw & \qw & \qw
    \end{quantikz}
    \caption{Representation of the corresponding circuit after a successful measurement outcome $\ket{0}$. This circuit is then repeated many times.}
    \label{fig:circuit}
\end{figure}
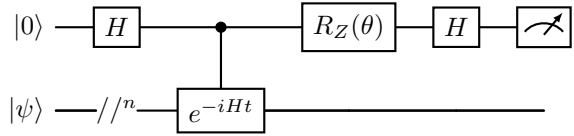

A single iteration of our approach is presented in Fig~\ref{fig:circuit}. The Hadamard test is applied to the evolution operator with a time parameter $t$ that can be adjusted. We also introduce a parametrized rotation $R_{Z}(\theta)$ over the Z axis, giving us another adjustable parameter, namely the rotation angle $\theta$. Note that while there is a single ancilla qubit, the quantum register for the current state $\ket{\psi}$ has an arbitrary number of qubits, which we denote $n$. 

Consider an initial state $\ket{\psi}$, for instance prepared in a uniform superposition in the computational basis by a Hadamard tower, followed by the evolution operator $U=e^{-iHt}$ controlled by the ancilla qubit. We express $\ket{\psi}$ in the energy eigenbasis of the cost Hamiltonian:

\begin{equation} 
\ket{\psi} = \sum_k c_k \ket{E_k}, 
\end{equation}

where $\ket{E_k}$ are the eigenstates of the Hamiltonian, and $E_0 < E_1 < \cdots < E_{\max}$ are their corresponding eigenvalues. The evolution operator $U$ can then be written as

\begin{equation} 
U = e^{-i H t} = \sum_k  e^{-i E_k t}\ket{E_k}\bra{E_k} . 
\end{equation}

Next, we apply the one qubit QPE, i.e., $U$ is applied conditionally on the state of an ancilla qubit initialized in $\ket{+}$, and then the ancilla is measured. The resulting joint state before measurement (after the last Hadamard gate in Figure~\ref{fig:circuit}) is

\begin{equation} 
\ket{\phi} = \ket{0} \frac{\ket{\psi} + e^{i\theta t} U \ket{\psi}}{2} + \ket{1} \frac{\ket{\psi} - e^{i\theta t} U \ket{\psi}}{2}. 
\end{equation}

The core idea is to adjust the parameters $t$ and $\theta$ such that a measurement of the ancilla in the $\ket{0}$ state will suppress amplitudes from excited states and amplify the overlap between the resulting state and the ground space. By adjusting the parameters of quantum interference, such as $(\theta, t)$, we can steer the quantum system to favor low-energy states, effectively acting as an energy filter that amplifies the probability of desirable solutions. Henceforth, outcome $0$ of the aancilla measurement will be denoted as the \emph{desired outcome} while outcome $1$ is the \emph{undesired} outcome. We will now describe our approach, focusing first on the case of the desired outcome and later on the undesired outcome.

\subsubsection{Desired outcome: suppress high-energy sector}
When the ancilla is measured in $\ket{0}$, the system is projected onto the following state (up to a normalization factor).

\begin{equation} 
\ket{\psi'_0} \propto \sum_k c_k e^{\frac{i(\theta - t E_k)}{2}} \cos\left( \frac{\theta - t E_k}{2} \right) \ket{E_k}. 
\end{equation}

We will now adjust the free parameters $t$ and $\theta$ to amplify the low-energy states and suppress the high-energy states. In order to perform this selection, our approach requires knowledge about the energy spectrum. More precisely, we need to be able to bound the energy spectrum.

\begin{figure}[htbp]
    \centering
    \includegraphics[height=0.75\linewidth]{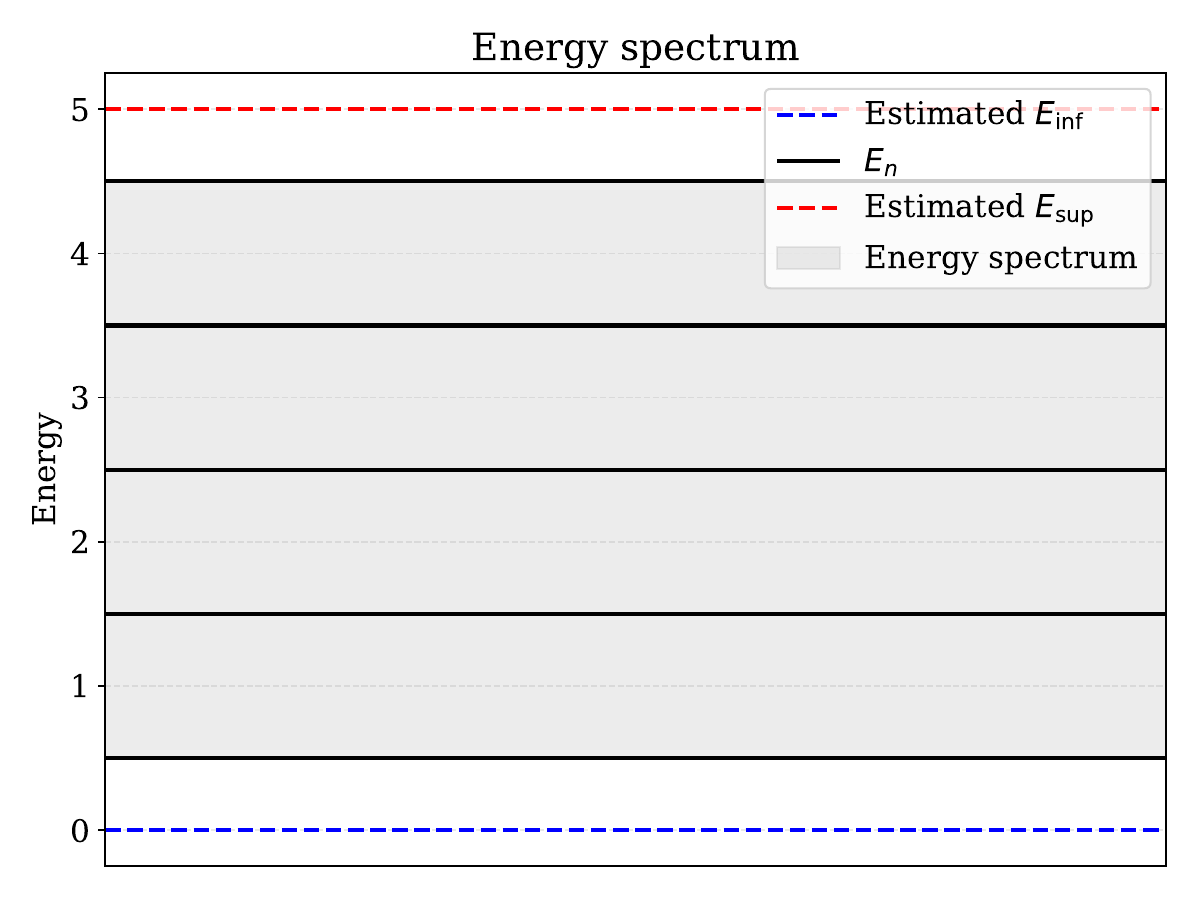}
    \caption{Energy spectrum illustrating a lower bound $E_{\inf}$ (corresponding to the ground state) and an upper bound $E_{\max}$ of the system’s energy. Quantum states closer to $E_0$ are favored by the post-selection mechanism, while those nearer to $E_{\max}$ are suppressed. This bounded energy distribution underlies the post-selection hypothesis used to amplify the ground-state contribution and attenuate higher-energy components.}
    \label{fig:energy_gap}
\end{figure}

\begin{hypothesis}
Both lower-bound, denoted $E_{\inf}$, and upper-bound, denoted $E_{sup}$, on the spectrum of the Hamiltonian are known. See Figure~\ref{fig:energy_gap} for a graphical representation.
\end{hypothesis}

Please note that we do not assume to know precisely the energy minimum $E_0$, nor the maximum energy $E_{\max}$. However, the tightness between the energy minimum $E_0$ and our bound $E_{\inf}$ will impact the convergence of the algorithm.

To enhance the amplitude of the ground state $\ket{E_0}$ while suppressing higher-energy components, we choose parameters $t$ and $\theta$ such that:

\begin{equation}
\cos \left(\frac{\theta - t E_{\inf}}{2}\right) = 1,
\end{equation}

\begin{equation}
    \cos\left(\frac{\theta - t E_{\sup}}{2}\right) = 0.
\end{equation}

As a result of measuring the ancilla qubit and obtaining $\ket{0}$, the amplitude of the ground state $\ket{E_0}$ increases in modulus and high energies are suppressed, according to

\begin{equation}\label{eq:cos}
c_k' = \frac{c_k\cos\vartheta_k} {\sqrt{\sum_j |c_j|^2  \cos^{2} \vartheta_j}}
\end{equation}

where

\begin{equation}
\vartheta_k = \frac{\pi}{2} \cdot \frac{E_k - E_{\inf}}{E_{\sup} - E_{\inf}}.
\end{equation}

In other words, the amplitude distribution is multiplied by a cosine function which is maximal for the lower-bound energy and zero for the upper-bound energy.

\begin{figure}[htbp]
    \centering
    \includegraphics[width=1\linewidth]{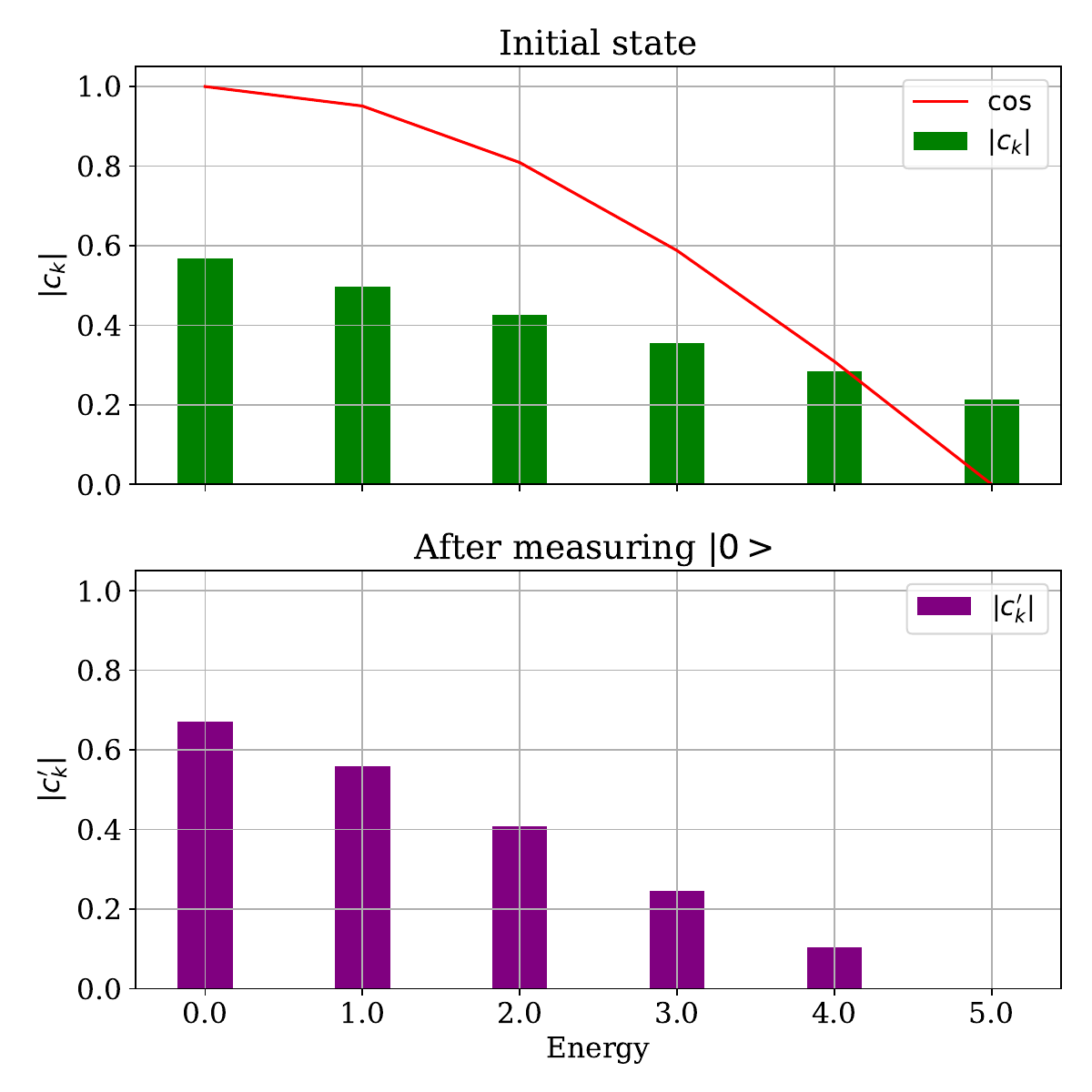}
    \caption{Energy distribution before and after measuring the QPE ancilla qubit in the desired outcome. The left image shows the state coefficients prior to QPE and the modulation by the cosine function. The right image displays the resulting state after QPE and obtaining the desired outcome, highlighting the amplification of low-energy components and suppression of high-energy components. For simplicity, we assumed that our energy bounds coincided with the minimal and maximal energies for this Figure.}
    \label{fig:cos}
\end{figure}

As illustrated in Figure~\ref{fig:cos}, the process exhibits amplification of low energy states, as described by the terms in Equation~\ref{eq:cos}.  The overlap with the ground state is amplified after each iteration since this coefficient is not suppressed by the cosine and due to renormalization. With a sufficient number of iterations, this leads to the ground state (corresponding to the minimum eigenvalue) becoming predominantly populated.

For simplicity, we assumed that our energy bounds coincided with the minimal and maximal energies for this Figure. In particular, having the upper-bound coincide with the maximal energy results in annihilation of the overlap with the highest-energy state. Should the bound not be tight, the highest-energy state will be suppressed with every iteration.

Thus, obtaining the desired outcome enhances the amplitude of the ground state while suppressing amplitudes from excited states, providing a quantum-native mechanism to bias the algorithm toward optimal solutions. 

\subsubsection{Undesired outcome: suppress low-energy sector}

Unfortunately, if the QPE ancilla is measured in the state $\ket{1}$, the resulting state has increased amplitudes over high-energy states, while the lower-energy components are suppressed, according to

\begin{equation}\label{eq:sin}
c_k' = \frac{c_k\sin\vartheta_k} {\sqrt{\sum_j |c_j|^2  \sin^{2} \vartheta_j}}
\end{equation}

where

\begin{equation}
\vartheta_k = \frac{\pi}{2} \cdot \frac{E_k - E_{\inf}}{E_{\sup} - E_{\inf}}.
\end{equation}

\begin{figure}[htbp]
    \centering
    \includegraphics[width=1\linewidth]{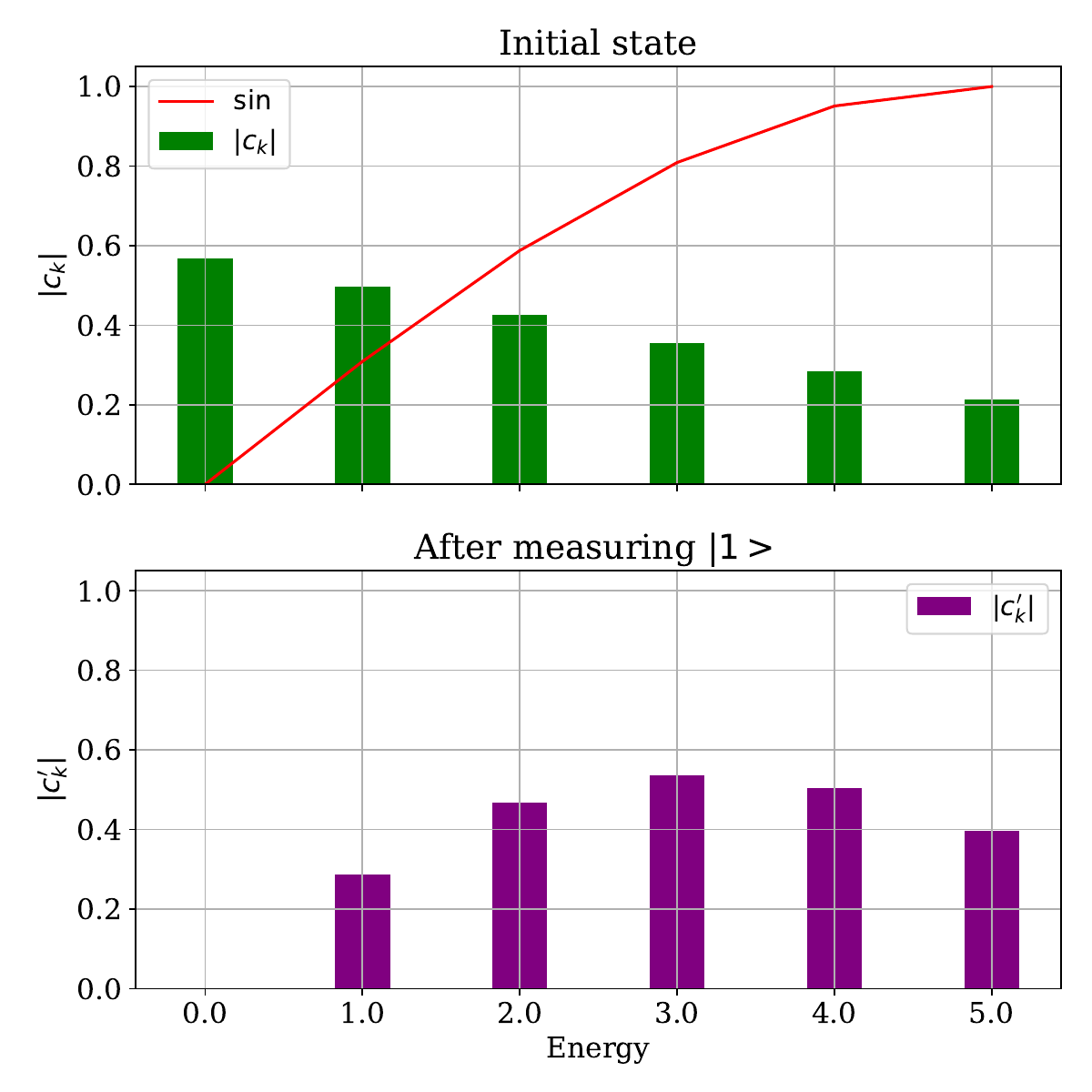}
    \caption{Energy distribution before and after measuring $\ket{1}$. The left image shows the initial state coefficients and the modulation by the sine function. The right image displays the resulting state after measuring the outcome $\ket{1}$, highlighting the amplification of high-energy components.}
    \label{fig:sin}
\end{figure}

As illustrated in Figure~\ref{fig:sin}, the process exhibits amplification of high energy states, as described by the terms in Equation~\ref{eq:sin}. 

Thus, our approach needs to be able to recover from an undesired outcome. Namely, it needs to recreate a state which has non-zero overlap with the ground space. Moreover, it should aim to transfer the amplitudes that are on the high-energy states back to the low-energy states.

\subsection{Repopulating low-energy eigenstates after a bad outcome}

Having demonstrated that measuring $\ket{0}$ on the ancilla qubit allows us to amplify the amplitude of the ground state, we now investigate how to repopulate the low-energy eigenstates after measuring $\ket{1}$ on the ancilla qubit. In other words, how can we design a fully coherent, unitary process that reproduces the same effect as measuring and selecting only the outcomes corresponding to the state \(\ket{0}\)?

To achieve this, we introduce a feedback mechanism: a heuristic based on the mixer strategy that can probabilistically steer the system back toward lower energy levels following an undesired measurement outcome. By analyzing the time evolution of Hamiltonians and examining the role of the mixer in the QAOA algorithm, we found that it is possible to bias the system toward the ground state using a fixed quantum operation. This approach is particularly effective for a certain class of Hamiltonians—specifically, combinatorial ones.

We modify the state by applying a rotation \(R_x(\sigma)\) to each qubit, with \(\sigma = \pi/2\) and $n=$number of qubits, resulting in a new unitary:
\begin{equation}
    U' = \left[\bigotimes_{k=0}^n R_x^{(k)}\left(\frac{\pi}{2}\right)\right] \cdot U
\end{equation}

It introduces a feedback dynamic: heuristically, $U'$ maps the high-energy sector to the low-energy sector. While it does not undo the measurement such as the approach of~\cite{Temme_Osborne_Vollbrecht_Poulin_Verstraete_2011}, it allows to recreate a state suitable for additional iterations. This coherent feedback mechanism offers a scalable and fully quantum approach to favoring the ground state, eliminating the need for classical post-processing and preserving the benefits of a fully unitary algorithm. Depending on the choice or construction of the Hamiltonian, the mixing angle may vary, and the standard X-mixer can be replaced by an XY-mixer or another type of mixer. 

\subsection{Incorporating Mid-Circuit Measurements}

With the two building blocks established; the one qubit QPE eigenvalue estimator and the feedback mechanism based on \( R_x(\pi/2) \). We now introduce the central feature of our approach: mid-circuit measurements.

Unlike their traditional use in error correction, we use mid-circuit measurements as an active part of the computation to steer the system toward its ground state. Our method embeds a one qubit QPE inspired routine into each QAOA layer using a single ancilla qubit, creating a dynamic feedback loop.

At each layer, the ancilla is initialized (see Figure~\ref{fig:circuit}) in the \(\ket{0}\) state and transformed by a Hadamard gate. We then apply a phase rotation \( R_z(\theta) \) on the ancilla with $\theta$ being a solution of Eq.~\ref{eq:12}, namely:

\begin{equation}
    \label{eq:12}
    \theta  = \frac{-\pi E_{\inf}}{E_{\sup} - E_{\inf}},
\end{equation}

where \( E_{\inf} \) and \( E_{\sup} \) bound the energy spectrum. This phase biases interference in favor of the ground state.

We then apply the evolution operator \( U = e^{-i H t} \), where the time parameter being a solution of Eq.~\ref{eq:13} , namely:

\begin{equation}
    \label{eq:13}
    t = \frac{-\pi}{E_{\sup} - E_{\inf}}.
\end{equation}

This evolution amplifies the energy-dependent phase differences between eigenstates. Crucially, the unitary operation is applied conditionally, controlled by the ancilla qubit. Finally, a Hadamard gate is applied to the ancilla prior to measurement.

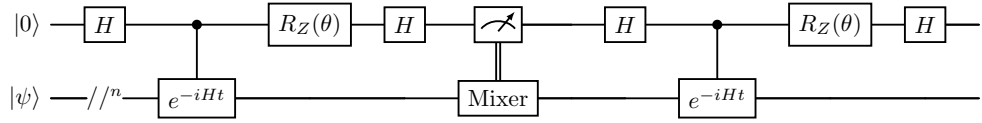
\begin{figure}[htbp]
    \centering
    \resizebox{\linewidth}{!}{%
    \begin{quantikz}
        \lstick{$\ket{0}$} & \gate{H} & \ctrl{1} & \gate{R_Z(\theta)} & \gate{H} & \meter{} \vcw{1} & \qw & \gate{H} & \ctrl{1} & \gate{R_Z(\theta)} & \gate{H} & \qw \\
        \lstick{$\ket{\psi}$} & \qw //^n & \gate[1]{e^{-iHt}} & \qw & \qw & \gate{\text{Mixer}} & \qw & \qw & \gate[1]{e^{-iHt}} & \qw & \qw & \qw
    \end{quantikz}
    }
    \caption{Two iterations of the circuit execution. Whenever the measurement outcome is $\ket{1}$, the mixer is applied. The measurement is assumed to reset the ancillary qubit to the $\ket{0}$ state before the next iteration.}
    \label{fig:circuit2}
\end{figure}

As we can see in Figure~\ref{fig:circuit2}, if the measurement outcome is \( 0 \), this indicates constructive interference towards with the ground state. If the outcome is \( 1 \), it indicates deviation, and we apply a mixer, for instance  \( R_x(\pi/2) \) to every system qubit. This reinjects superposition and re-explores low-energy configurations, effectively compensating for the unwanted projection without destroying coherence.

This process repeats across layers, creating a coherent energy amplification mechanism driven by mid-circuit measurements. The complete algorithm is presented below:\\

\begin{algorithmic}[1]\label{pseudocode}
\State Initialize system in state \( \ket{+}^{\otimes n} \otimes \ket{0}_\text{ancilla} \)
\For{each layer}
    \State Apply Hadamard gate on the ancilla qubit
    \State Apply \( R_z(\theta) \) on ancilla, with \( \theta = \frac{-\pi E_0}{E_{\max} - E_0} \)
    \State Apply controlled-\( U = e^{-i H t} \) with \( t = \frac{-\pi}{E_{\max} - E_0} \)
    \State Apply Hadamard gate on the ancilla qubit
    \State Measure ancilla qubit
    \If{ancilla = 1}
        \State Apply \( R_x(\pi/2) \) to each system qubit
    \EndIf
\EndFor
\State Measure all system qubits
\end{algorithmic}

\subsection{Problem Definition and Experimental Setup}\label{subsec:problem_definition}

Having fully defined our algorithm, we proceeded to evaluate its performance through numerical simulations. For our experiments, we primarily used the standard QAOA framework applied to the \textit{vertex cover} problem, formulated as a QUBO Hamiltonian~\cite{Lucas_2014}.

By carefully selecting the parameters, this formulation guarantees that the lowest energy configuration corresponds to a minimal vertex cover, with energy exactly zero when all constraints are satisfied. By shifting the energy by -1, the QUBO encodes both correctness and optimality in its energy landscape, making it suitable for our approach.

In addition to this main case study, we explored our method on a variety of other optimization problems. While the vertex cover served as our core benchmark, it is important to note that any NP-complete problems can be mapped to another with polynomial overhead. Thus, improvements demonstrated on this instance are expected to generalize across a broader class of combinatorial problems.

\subsubsection{Minimum Vertex Cover}

Given an undirected graph $G = (V, E)$, a vertex cover is a subset $C \subseteq V$ such that every edge $(u, v) \in E$ has at least one endpoint in $C$. Formally,

\begin{equation}\label{eq:4}
C \subseteq V, \quad 
\text{such that } 
\begin{aligned}[t]
& \forall (u, v) \in E, \\
& \quad u \in C \quad \text{or} \quad v \in C
\end{aligned}
\end{equation}

The objective is to find the smallest such set $C$.

\subsubsection{Formulating Vertex Cover as a QUBO}

The minimum vertex cover problem can be naturally expressed as a Quadratic Unconstrained Binary Optimization (QUBO) problem, which can then be mapped onto a quantum Hamiltonian using Pauli operators\cite{Lucas_2014}.

Given a graph $G = (V, E)$ with $|V| = n$, we define binary variables:

\begin{equation}\label{eq:5}
x_i =
\begin{cases} 
1, & \text{\shortstack{if vertex $i$ \\ is included in the vertex cover}}, \\
-1, & \text{otherwise}.
\end{cases}
\end{equation}

The objective function aims to minimize the number of selected vertices:

\begin{equation}\label{eq:6}
\min A \sum_{i \in V} x_i.
\end{equation}

To enforce the constraint that each edge $(u, v) \in E$ has at least one of its endpoints in the vertex cover, we introduce a penalty term:

\begin{equation}\label{eq:7}
H_{\text{penalty}} = B \sum_{(u,v) \in E} (1 - x_u)(1 - x_v).
\end{equation}

The full QUBO Hamiltonian is given by:

\begin{equation}\label{eq:8}
H = A \sum_{i \in V} x_i + B \sum_{(u,v) \in E} (1 - x_u)(1 - x_v),
\end{equation}

where $A < B$, and $B$ is a sufficiently large penalty weight to ensure that the constraints are satisfied.

\subsubsection{Applying the Theory to a Small Graph}

\begin{figure}[htbp]
    \centering
    \includegraphics[width=0.5\linewidth]{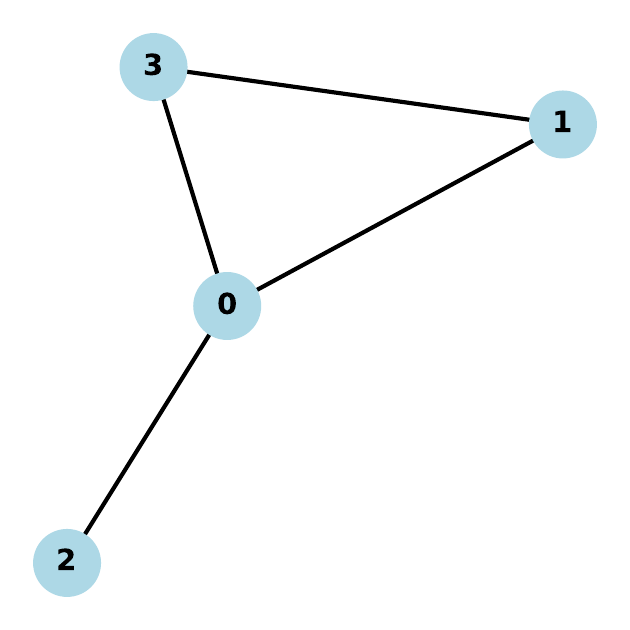}
    \caption{Graph structure used to encode the problem instance.}
    \label{fig:4_graph_nodes}
\end{figure}

Consider the undirected graph $G = (V, E)$ shown in Fig.~\ref{fig:4_graph_nodes}, with
\[
V = \{0, 1, 2, 3\}, \quad
E = \{(0,1), (0,2), (0,3), (1,3)\}.
\]
The degrees of the vertices are $d_0 = 3$, $d_1 = 2$, $d_2 = 1$, and $d_3 = 2$.

Following Eq.~\eqref{eq:8}, the QUBO Hamiltonian in the $\{\pm 1\}$ encoding is
\begin{equation}
H = A\sum_{i \in V} x_i + B\sum_{(u,v) \in E} (1 - x_u)(1 - x_v), \quad A < B,
\end{equation}
where $x_i = +1$ if vertex $i$ is included in the cover, and $x_i = -1$ otherwise.

Expanding the penalty term gives
\begin{align}
H &= B|E| + \sum_{i \in V} \left(A - B\,d_i \right) x_i
    + B \sum_{(u,v) \in E} x_u x_v, \label{eq:qubo_expanded}
\end{align}
where $|E| = 4$. Substituting the vertex degrees for this graph, we obtain
\begin{align}
H &= 4B 
    + (A - 3B)x_0 
    + (A - 2B)x_1 \nonumber\\
  &\quad + (A - B)x_2 
    + (A - 2B)x_3 \nonumber\\
  &\quad + B \bigl( x_0 x_1 + x_0 x_2 \nonumber\\
  &\quad\quad + x_0 x_3 + x_1 x_3 \bigr).
\end{align}

For concreteness, choosing $A = 1$ and $B = 2$ (which satisfies $B > A/2$) yields
\begin{align}
H &= 8 - 5x_0 - 3x_1 - x_2 - 3x_3 \nonumber\\
&\quad + 2\left( x_0x_1 + x_0x_2 + x_0x_3 + x_1x_3 \right).
\end{align}

\subsubsection{Minimum Vertex Covers}
The optimal solutions to this QUBO correspond to the smallest vertex covers of $G$.  
In this example, all minimum covers have size $|C| = 2$, namely:
\[
C = \{0, 1\} \quad \text{and} \quad C = \{0, 3\},
\]
corresponding to configurations $x = (+1, +1, -1, -1)$ and $x = (+1, -1, -1, +1)$, respectively.  
No vertex cover of size~1 exists for this graph.

\subsubsection{Pauli Operator Form}
Mapping $x_i \mapsto Z_i$ (Pauli $Z$ operator), Eq.~\eqref{eq:qubo_expanded} becomes the Ising Hamiltonian
\begin{equation}
\hat{H} = B|E|\,{I} + \sum_{i \in V} (A - B\,d_i) Z_i
          + B \sum_{(u,v) \in E} Z_u Z_v,
\end{equation}
up to an additive constant $B|E|$, which is irrelevant for optimization.  
This form can be directly implemented as the Hamiltonian in our quantum optimization algorithms.

\subsubsection{Choosing $A$ and $B$ for Analytical Estimation of $E_{\inf}$ and $E_{\sup}$}
We choose $A$ and $B$ such that the extremal energies of the QUBO Hamiltonian can be estimated directly from the problem parameters, without explicit diagonalization. 

Starting from the Ising formulation in Eq.~\eqref{eq:qubo_expanded}, with $d_i$ denoting the degree of vertex $i$ and $x_i \in \{\pm 1\}$ representing the spin variables, we can derive the relevant energy bounds as follows:
\begin{itemize}
    \item Choose $B$ as a simple multiple of $A$, e.g., $B = 2A$, to ensure $B > A/2$ (constraint satisfaction) while keeping coefficients integer-valued.
    \item The maximum energy $E_{\max{}}$ occurs when all $x_i = -1$ (no vertex in the cover), which maximizes all penalty terms:
    \[
    E_{\max{}} = A(-n) + B \cdot 4|E| \quad \text{(in $\pm 1$ encoding)}.
    \]
    \item The minimum energy $E_{0}$ occurs for a valid minimum vertex cover $C_{\min}$ of size $k_{\min}$, with $x_i = +1$ for $i \in C_{\min}$ and $x_i = -1$ otherwise:
    \[
    E_{0} = A(2k_{\min} - n),
    \]
    since all penalties vanish in a perfect cover.
\end{itemize}

For our example with $n = 4$, $|E| = 4$ where $n=$ numbers of nodes, and $k_{\min} = 2$, taking $A=1$ and $B=2$ gives:
\[
E_{0} = 2\cdot 2 - 4 = 0, 
\qquad 
E_{\max{}} = -4 + 2 \cdot 4 \cdot 2 = 12.
\]
These values are obtained without diagonalization, using only $n$, $|E|$, and $k_{\min}$.

\section{Results}
\begin{figure}[htbp]
    \centering
    \includegraphics[height=1\linewidth]{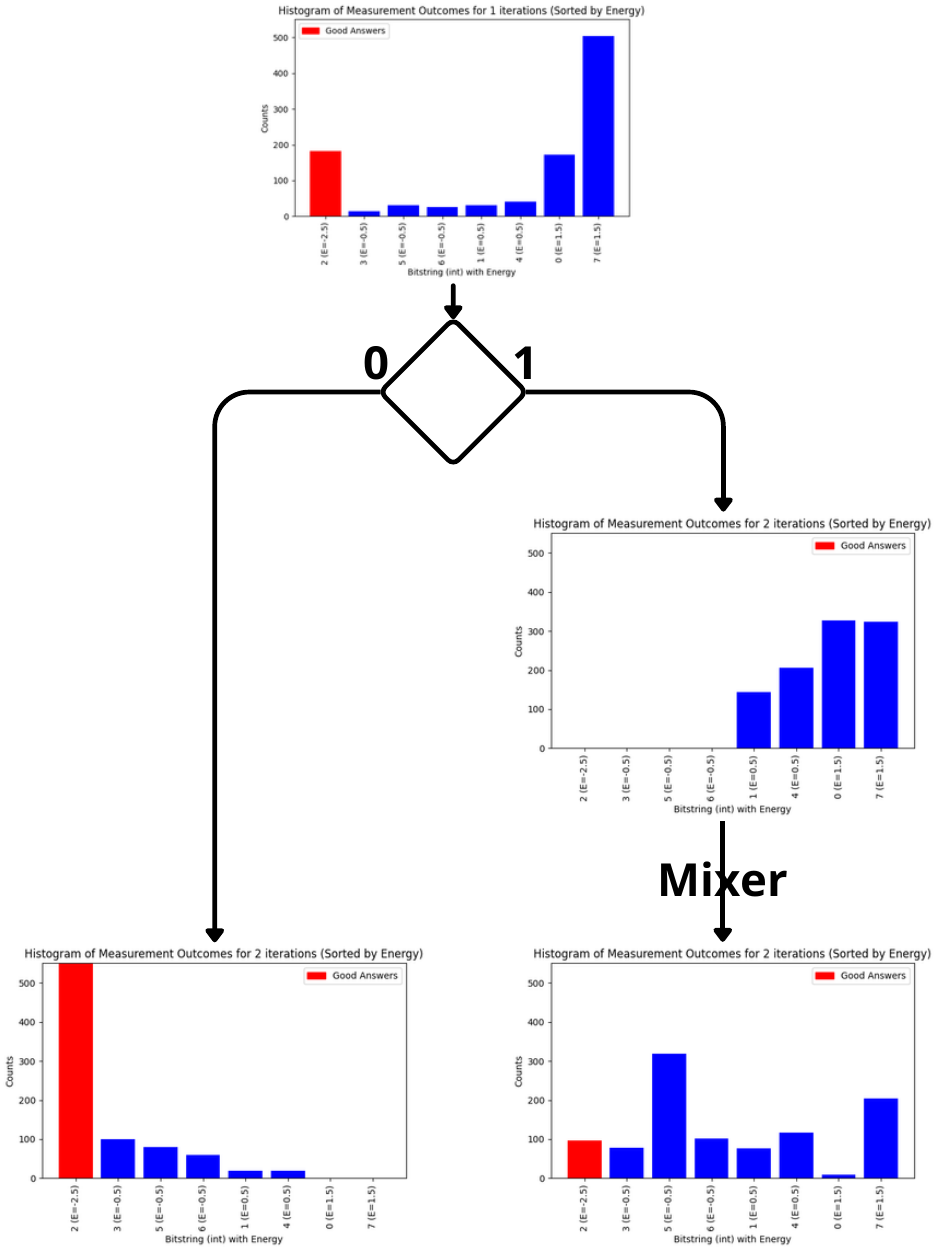}
    \caption{Comparison of state evolution based on ancilla measurement. The red bar represents the optimal solution. The left part shows the resulting state after obtaining the desired outcome upon measuring the QPE ancilla qubit. The right part shows the resulting state after obtaining the undesired outcome and then after applying the mixer.}
    \label{fig:path_outcome}
\end{figure}

\subsection{Ancilla Measurement and State Evolution}

To provide intuition about the algorithm's internal dynamics, we examine a small instance of the Vertex Cover problem and track the quantum state across a single iteration, as illustrated in Figure~\ref{fig:path_outcome}. The figure begins with an arbitrary superposition representing the state of the system after previous iterations. We then analyze how the system evolves depending on the outcome of measuring the ancilla qubit.

The desired outcome path corresponds to the left part (measured $\ket{0}$) of Figure~\ref{fig:path_outcome}. The resulting quantum state has significantly increased overlap with the (unique) ground state. Moreover, since we assume a tight upper bound on energy, the overlap over the (degenerate) highest-energy subspace is annihilated. This is perfectly in accordance with theory. 

The undesired outcome path corresponds to the right part (measured $\ket{1}$) of Figure~\ref{fig:path_outcome}. The resulting quantum state has now zero overlap with the (unique) ground state and the higher energy states were favored. In this less favorable case, the mixer plays a critical role in repairing the state. By redistributing amplitude through quantum interference, the mixer partially recovers overlap with the target solution. As seen in the bottom part of the figure, after applying the mixer, the state shows non-zero amplitude on the red bar, which corresponds to the optimal solution. 

This example illustrates the delicate balance between measurement-induced collapse and the corrective power of mixing operations. It highlights how the algorithm leverages quantum interference not only to reinforce correct paths but also to recover from misaligned projections. Such behavior is especially important when scaling to larger problem instances.

\subsection{Convergence Behavior and Probability of Measuring the Optimal Solution}

Using the algorithm described in Pseudocode~\ref{pseudocode}, we define the probability of measuring the optimal solution(s) as a way to estimate how effectively the system evolves toward the desired outcome. This quantity is calculated as:

\begin{equation}
\begin{split}
P_{\text{good}} &= \sum_k \Pr(\ket{\psi_k}), \\
\text{where } \ket{\psi_k} &\text{ is a Good Answer}.
\end{split}
\end{equation}

This value is estimated by likelihood of sampling an optimal solution from the quantum state, normalized over the total number of shots. It serves as a key performance indicator for the algorithm's ability to converge toward the ground state.

\begin{figure*}[t]
    \centering 
    \includegraphics[height=0.75\linewidth]{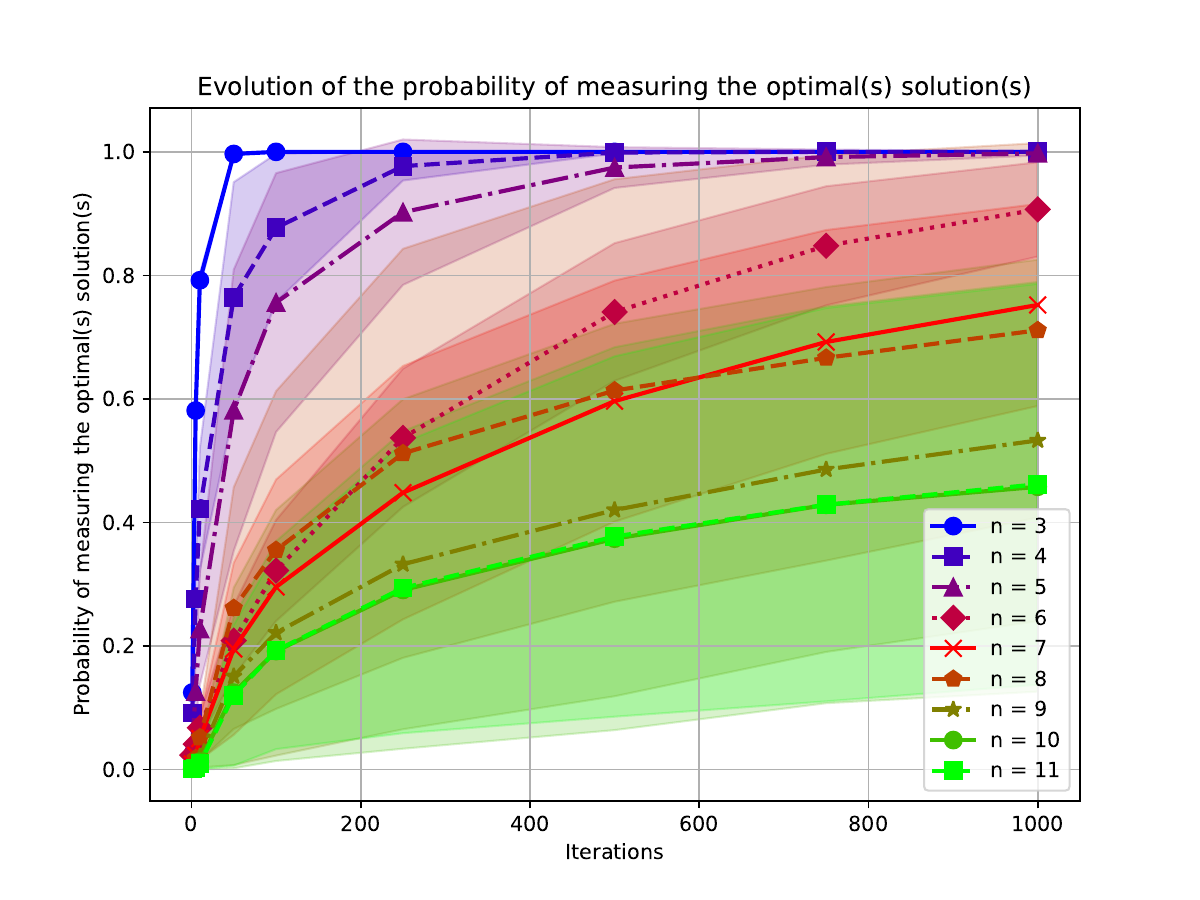}
    \caption{Exponential convergence behavior of the algorithm. In this experiment, we generate random graphs with a specific number of nodes, denoted by $n$, which also represents the number of qubits required to encode the graph.}
    \label{fig:evolution_of_pseudo_aproximation_ratio}
\end{figure*}

As shown in Figure~\ref{fig:evolution_of_pseudo_aproximation_ratio}, the system exhibits exponential convergence behavior. We observe that as the number of qubits increases, the number of iterations required for convergence also grows. Furthermore, the convergence rate slows for larger problem instances. This is expected, as the Hilbert space expands exponentially with the number of qubits, increasing the difficulty of amplifying the correct solution without more iterations or improved interference strategies.

An important consideration is the scaling of the variance as the problem size increases. As the number of nodes grows, the randomized graphs become more diverse. Larger graphs generate exponentially more candidate solutions, which dilutes the probability of sampling any single optimal configuration. When the problem admits only one unique optimal solution, the probability of observing it remains small, leading to higher statistical fluctuations and less precise estimates. For instance, in our experiments we observed cases where the probability of the correct solution could be around $20\%$, while the remaining probability mass was distributed across hundreds of suboptimal outcomes. In contrast, when the problem admits multiple optimal solutions, their probabilities accumulate and evolve coherently during the algorithm, effectively increasing $P_{\mbox{good}}$ and reducing variance. Thus, instances with multiple optima are easier to detect reliably, while unique-solution cases are inherently noisier and require more samples for accurate estimation.

Finally, while the observed exponential convergence is encouraging, it should not be misinterpreted as a pathway to solving NP-hard problems in polynomial time. The exponential growth of the Hilbert space guarantees that the number of required iterations, circuit depth, or sampling overhead will still scale unfavorably with problem size. Thus, the observed convergence rate demonstrates the efficiency of interference-based amplification for moderate system sizes, but does not circumvent the fundamental complexity-theoretic barriers of NP-hard optimization.

\subsection{Comparing VQA to our Approach on quantum simulators}

Our method is inspired by Quantum Phase Estimation (QPE), but it leverages quantum interference in a way that is more akin to Grover’s algorithm~\cite{Grover_1996} coherently amplifying the amplitude of good solutions through repeated constructive interference. We chose to compare our approach with the Quantum Approximate Optimization Algorithm (QAOA) due to the structural similarities in the ansatz used in both algorithms.

QAOA relies on a variational strategy that involves an \textit{outer classical optimization loop} to iteratively refine parameters. In contrast, our approach uses fixed operations with no classical feedback, and instead harnesses interference and mid-circuit measurements to guide the system toward optimal states.

In QAOA, the total computational cost is determined by both the number of classical optimization steps and the number of quantum measurements (shots) per step. This can be expressed as:

\begin{equation}
\text{Total shots (QAOA)} = w \times l
\end{equation}

where $w$ is the number of outer optimization iterations, and $l$ is the number of shots per iteration. Our method eliminates the outer loop entirely and requires only $l$ shots, reducing both classical overhead and the need for repeated circuit recompilation and parameter reloading.

Another important distinction lies in circuit depth. QAOA circuits are typically shallow especially for small depth values ($p$) making them appealing for near-term devices. Our circuits are deeper due to repeated controlled unitaries, akin to multiple rounds of phase amplification. 


To evaluate performance, we compared both methods on a small instance of the Vertex Cover problem. QAOA was run with depth $p=5$ and 50 gradient descent steps. Our method used a circuit with equivalent total depth (50 interference steps) but required no classical optimization.

\begin{figure}[h]
    \centering
    \includegraphics[width=0.65\linewidth]{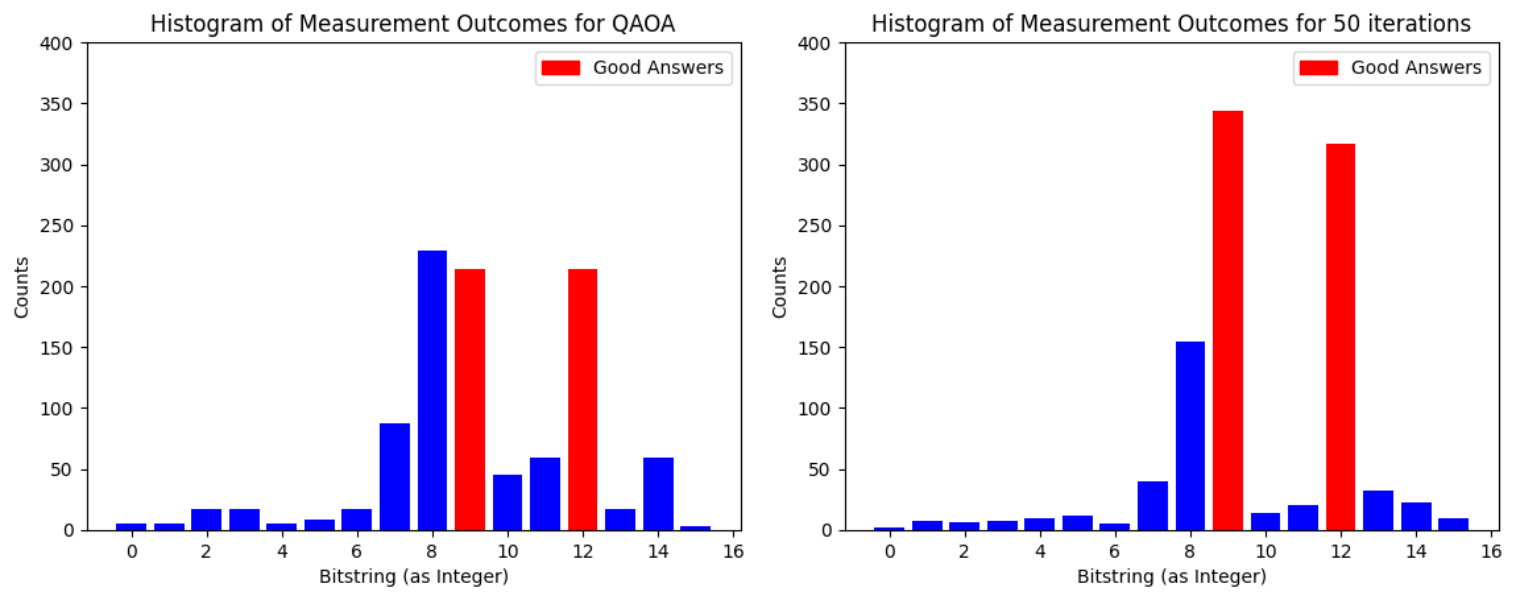}
    \caption{Comparison of solution quality between QAOA (depth $p=5$, 50 optimization steps) and our interference-based approach (depth = 50, no classical optimization). Our method shows competitive performance while avoiding the overhead of classical optimization, suggesting potential advantages in scalability and robustness.}
    \label{fig:comparison_qaoa}
\end{figure}

The results presented in Figure~\ref{fig:comparison_qaoa} suggest that while our approach involves deeper circuits, it maintains high solution quality without relying on classical parameter tuning. This highlights its potential for applications on near-term quantum hardware, particularly where circuit recompilation and hybrid feedback loops pose implementation challenges.

\begin{figure}[h]
    \centering
    \includegraphics[width=0.65\linewidth]{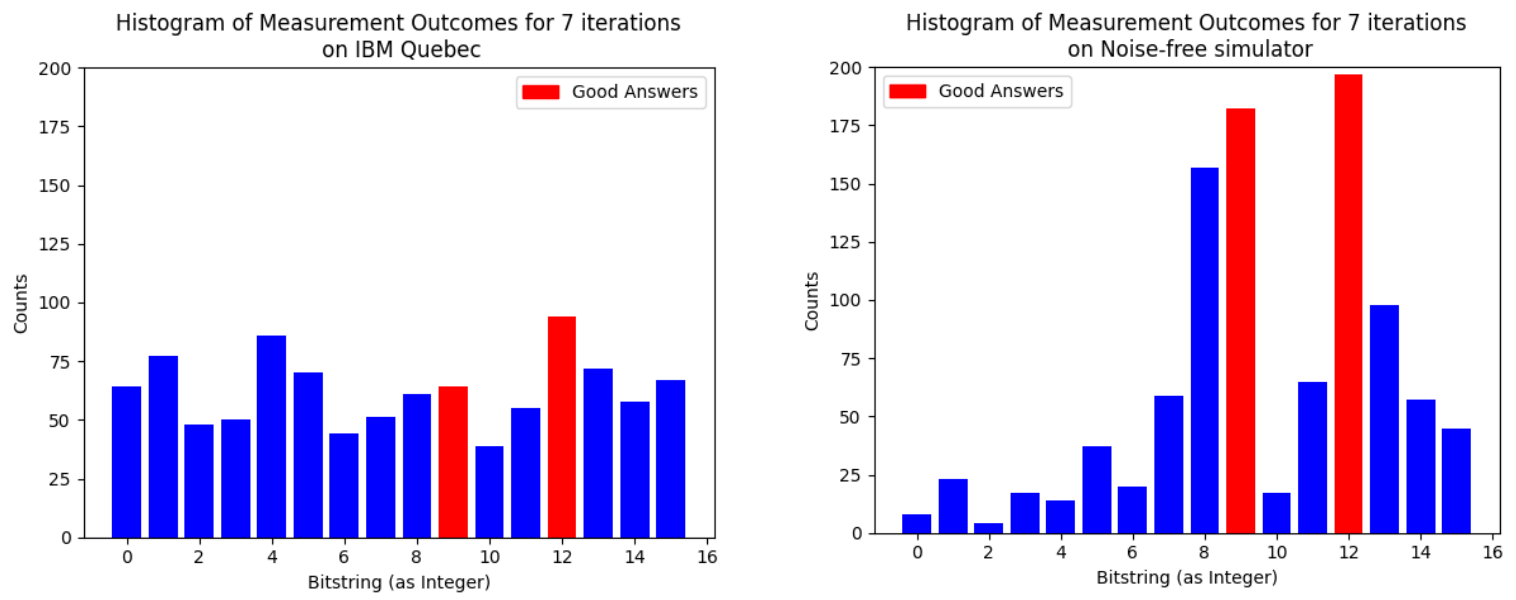}
    \caption{Execution of our algorithm on real quantum hardware for a 4 nodes graph over 7 iterations. While full convergence is not achieved, trends begin to emerge that favor low-energy solutions.}
    \label{fig:execution_real_hardware}
\end{figure}

\subsection{Execution on Real Hardware}

To evaluate the feasibility of our approach on existing quantum devices, we deployed our algorithm on IBM’s superconducting quantum hardware. Figure~\ref{fig:execution_real_hardware} shows the outcome of executing our circuit on a 4-node graph from Fig.~\ref{fig:4_graph_nodes} over the course of 7 iterations.


As expected, there is a noticeable discrepancy between the ideal simulation and the results obtained from real hardware. This performance gap arises primarily from two factors: hardware noise and limited qubit connectivity. To satisfy the connectivity constraints of IBM’s architecture, the compiler introduces multiple SWAP gates, which increase circuit depth and significantly degrades fidelity.

Moreover, the use of Toffoli gates and other multi-qubit controlled operations presents a major challenge on current NISQ hardware. These gates are decomposed into a series of two-qubit and single-qubit gates during transpilation, further increasing circuit depth and making the execution more susceptible to noise.

This practical experiment highlights current hardware limitations in executing deep and entangled circuits—particularly those involving repeated ancilla interactions and mid-circuit measurements. However, it also motivates an important open question: whether targeted error correction or mitigation, applied specifically to the ancilla qubit, could serve as a low-overhead strategy to improve performance. Such techniques could provide a scalable path forward for hybrid algorithms that rely on interference and measurement-driven feedback.

\section{Discussion}

\subsection{The Crucial Role of the Ancilla Qubit}
\begin{figure}[h]
    \centering
    \includegraphics[width=0.65\linewidth]{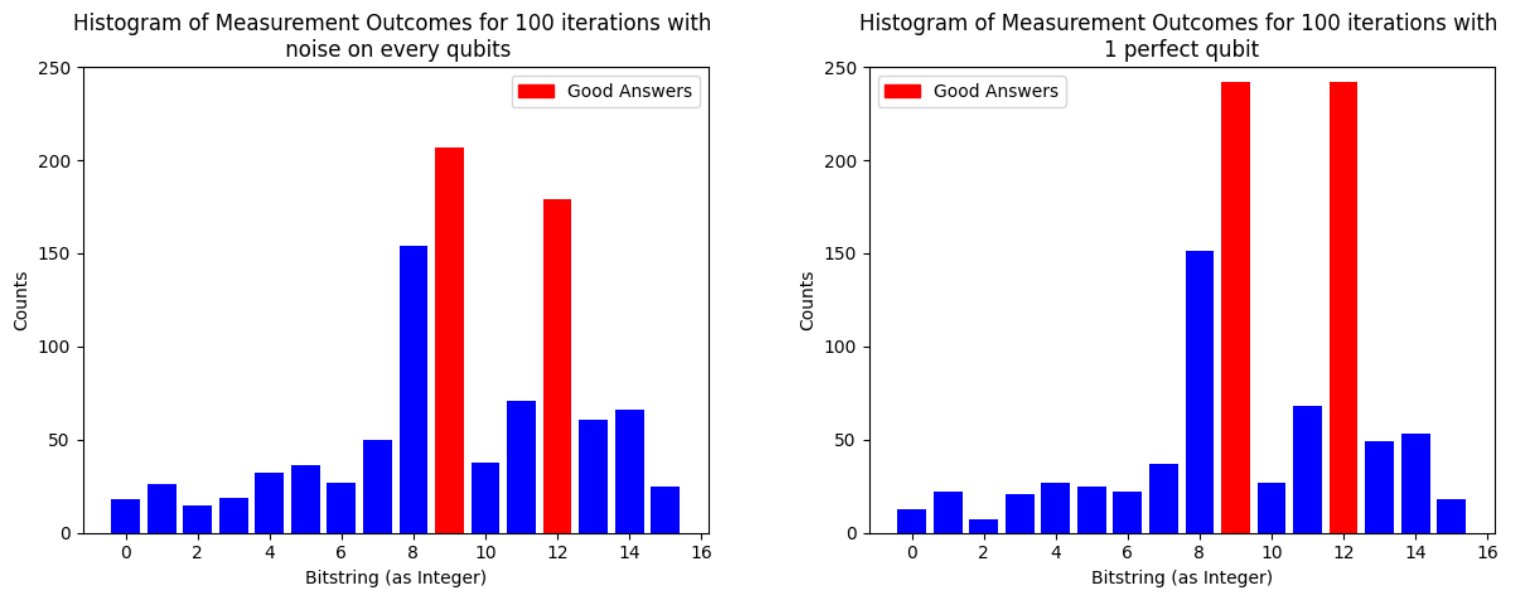}
    \caption{Comparison of performance between a fully noisy system (FakeQuebec~\cite{IBMQuantum}) and one with a noise-free ancilla qubit on a small graph instance. The presence of a protected ancilla substantially improves success probability, illustrating its central role in guiding the system’s evolution toward optimal solutions.}
    \label{fig:comparison_between_noise_model}
\end{figure}

Our results underscore the pivotal role played by the ancilla qubit in enabling the feedback mechanism central to our algorithm’s success. The simulations comparing scenarios with a noise-free ancilla against fully noisy systems (Figure~\ref{fig:comparison_between_noise_model}) reveal a striking difference: selectively protecting the ancilla significantly improves the probability of sampling optimal solutions. This suggests that the ancilla qubit effectively functions as a “quantum control hub,” driving constructive interference that steers the system toward the ground state.

This behavior can be interpreted as a form of implicit error filtering. When the ancilla measurement successfully projects the system closer to the ground state, constructive interference amplifies the desired solution component. Even if perfect amplification does not occur at every step, the iterative feedback gently rotates the state toward low-energy eigenstates. Conversely, noise on the ancilla disrupts this delicate interference pattern, quickly degrading the algorithm’s efficacy and potentially propagating errors throughout the system.

These insights have profound implications for near-term quantum hardware and algorithm design. They highlight the importance of \textit{targeted error mitigation or correction focused on the ancilla qubit}, rather than uniform protection across all qubits. Given the resource constraints on current devices, prioritizing the ancilla for error correction could be a pragmatic way to enhance algorithmic performance without the overhead of full error correction on all qubits.

While we leave the detailed exploration of ancilla-focused error correction to future work, this direction is promising. If achievable, it would enable robust, scalable implementations of our interference-based algorithm on NISQ devices. Such a capability could represent a crucial step toward realizing utility-scale quantum optimization by stabilizing the feedback loop at the heart of our method.

In summary, our findings emphasize that the ancilla qubit is not merely a passive participant but a critical driver of algorithmic success. Ensuring its coherence and reliability may unlock new pathways for near-term quantum algorithms that rely on measurement-based feedback and interference amplification.

\subsection{Mid-Circuit Measurement as an Algorithmic Primitive}

While mid-circuit measurement has predominantly been investigated in the context of quantum error correction, our results highlight its broader significance as a powerful algorithmic primitive. In our approach, repeated mid-circuit measurements of the ancilla qubit enable a dynamic feedback mechanism that guides the quantum state towards optimal solutions through constructive interference.

This use of mid-circuit measurement extends beyond mere error detection and correction it provides a real-time control tool within the algorithm itself, allowing for iterative state refinement without the overhead of classical optimization loops. By demonstrating the practical integration of mid-circuit measurement and feedback in an optimization algorithm, our work illustrates that this primitive can be harnessed to implement measurement-driven quantum computations that actively steer amplitude amplification and solution sampling.

Looking ahead, mid-circuit measurement is poised to play a central role in next-generation quantum algorithms, particularly on near-term devices where circuit depth and noise remain limiting factors. Future research could explore hybrid strategies combining mid-circuit measurement feedback with variational parameter tuning, potentially improving convergence rates and robustness.

Moreover, advancements in quantum hardware—such as faster measurement protocols, low-latency classical feedforward, and efficient qubit resets will be critical to fully realizing the potential of mid-circuit measurement as an algorithmic primitive. Our findings lay foundational evidence that mid-circuit measurements are not only a cornerstone for error correction but can also enable novel algorithmic structures, expanding the landscape of quantum computation beyond traditional paradigms.

\section{Conclusion}

In this work, we introduced a novel quantum optimization approach that leverages mid-circuit measurements and coherent feedback within a single-qubit variant of QPE. This technique amplifies the ground state amplitude of a problem Hamiltonian by harnessing quantum interference and partial error resilience, all without relying on classical optimization loops.

Unlike conventional variational algorithms such as QAOA, our method circumvents the need for costly outer optimization routines, instead exploiting intrinsic quantum dynamics to progressively steer the system toward low-energy solutions. Through simulations, we demonstrated exponential convergence to optimal states and showed competitive performance compared to QAOA with fewer circuit executions. Additionally, we underscored the pivotal role of the ancilla qubit, as a driver of constructive interference.

Experimental runs on real quantum hardware validated the feasibility of our algorithm in practice, while also revealing the challenges imposed by hardware noise, limited connectivity, and the overhead of multi-qubit gate decompositions. These findings highlight important hardware considerations for implementing feedback-driven quantum algorithms.

Our results pave the way for a new class of hybrid quantum algorithms that fully exploit mid-circuit measurement and real-time feedback as fundamental algorithmic primitives. Future research directions include the development of targeted error correction or mitigation strategies specifically for the ancilla qubit, exploration of more sophisticated feedback mechanisms, and scaling studies on larger problem instances and emerging quantum architectures.

By demonstrating that measurement-based feedback can be integrated effectively beyond error correction, this work contributes to broadening the landscape of quantum optimization and opens promising avenues toward practical quantum advantage in NISQ devices and beyond.

\section*{Acknowledgement}
We would like to thank the Plateforme d’Innovation Numerique et Quantique du Québec (PINQ2) for the access to the machine ibm\_quebec and the computation time needed for this study. This work was supported by Mitacs/PINQ² under project IT42780. We would like to thank Juliette Geoffrion from Calcul Qu\'ebec for her careful revision and insightful comments on this manuscript.

\pagebreak

\noindent
\bigskip

\bibliography{sn-bibliography}

\end{document}